# Neuromorphic Computing for Low-Power Artificial Intelligence

Keshava Katti[1], Pratik Chaudhari[1], Deep Jariwala[1].

[1] Electrical & Systems Engineering, University of Pennsylvania, Philadelphia, USA.

*"Heightened demand for computing is creating new risks because global energy production cannot keep up, but neuromorphic computing offers an energy-efficient path forward."*

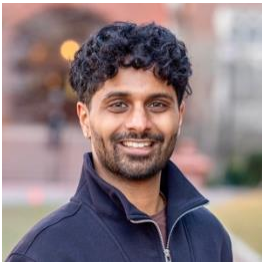
Keshava Katti

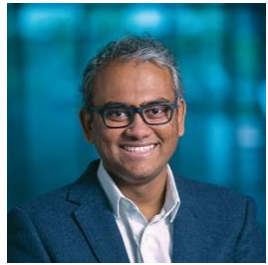
Pratik Chaudhari

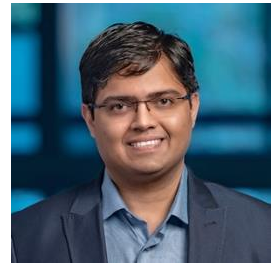
Deep Jariwala

Classical computing is beginning to encounter fundamental limits of energy efficiency. This presents a challenge that can no longer be solved by strategies such as increasing circuit density or refining standard semiconductor processes. The growing computational and memory demands of artificial intelligence (AI) require disruptive innovation in how information is represented, stored, communicated, and processed. By leveraging novel device modalities and compute-in-memory (CIM), in addition to analog dynamics and sparse communication inspired by the brain, neuromorphic computing offers a promising path toward improvements in the energy efficiency and scalability of current AI systems. But realizing this potential is not a matter of replacing one chip with another; rather, it requires a co-design effort, spanning new materials and non-volatile device structures, novel mixed-signal circuits and architectures, and learning algorithms tailored to the physics of these substrates. This article surveys the key limitations of classical complementary metal-oxide-semiconductor (CMOS) technology and outlines how such cross-layer neuromorphic approaches may overcome them.

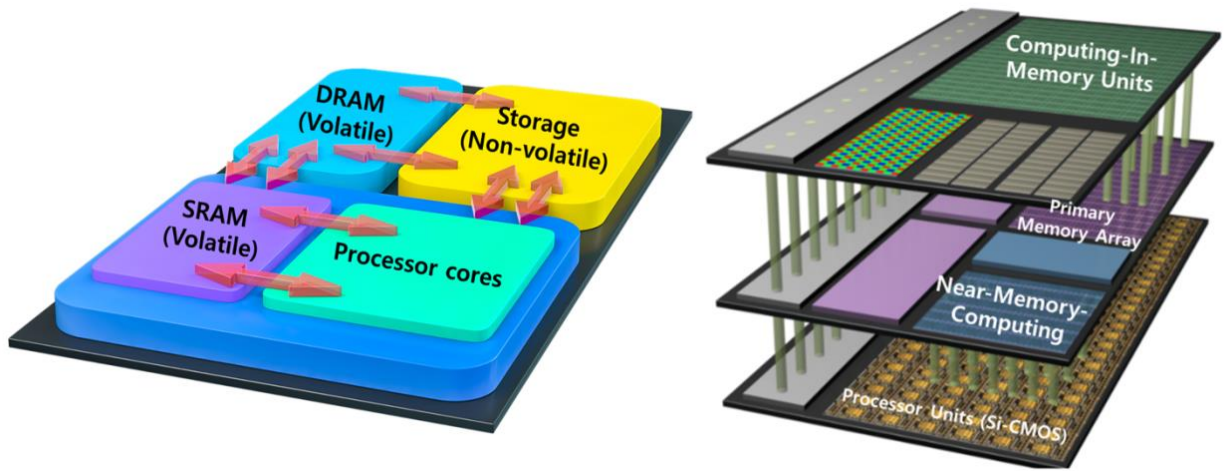

**Figure 1:** Comparison of classical and neuromorphic computer architectures. Source: (Kim, Karpov, et al. 2023).

**Current Challenges to Classical Computing**

Current scaling trends of AI hardware are becoming prohibitive with respect to power consumption and cost. Specifically, communication cost is associated with existing von Neumann architecture, which represents any stored-program computer for which the fetch instruction and data operation cannot occur simultaneously because they share a common bus (Liu et al. 2023). In most implementations today, dynamic random-access memory (DRAM), static random-access memory (SRAM), and flash storage are placed separately from processor cores in a chip layout on the millimeter to centimeter scale, as shown in Figure 1 (Left), incurring both time and energy costs for transfer of data (Kim, Karpov, et al. 2023). It is projected that the total global energy dedicated to communicating bits of information will reach $10^{20}$ J/year by around 2035, overtaking the total energy dedicated to computing bits of information and approaching the world's total generated energy at about $10^{21}$ J/year only a decade after that (Decadal Plan for Semiconductors - SRC).

In contrast, cortical gray matter in the human brain consumes about 3 W of power out of the 17–20 W of available glucose; even after 3× this budget (~ 9 W) is lost to heat, the human brain still outperforms classical computing by nearly an order of magnitude, with gray matter using about 2 nW per neuron (Balasubramanian 2021). Within gray matter's 3 W budget, over half (~ 1.67 W) is dedicated to communication, 35× less is used for computation (~ 0.05 W), and the rest is allocated to synaptic modification (Levy and Calvert 2021). Indeed, the human brain dedicates a surprising amount of its available glucose to communication, but this allocation remains feasible due to its ultra-low per-neuron power, which sums both computation and communication cost of gray matter and normalizes by the several billion neurons comprising this kind of brain tissue. One promising bio-inspired strategy encodes information using sparse, asynchronous spikes that offer a combinatorially large information capacity, utilizing a simple set of look-up tables to emulate recurrent neural network (RNN) and transformer models in a spiking format with a 10,000-fold reduction in computational resources over non-spiking counterparts (Izhikevich 2025). The next commercially-viable hardware platform for AI must meet such top-down algorithmic approaches with bottom-up hardware that further reduces energy scaling in the number of computational units, and the best place to start is by driving down communication cost.

We can better understand this power consumption problem by noting that the rate of increase in large language models (LLMs) at 240 times per 2 years in transformer size is far outpacing the 2 times per 2 years rate of increase in the underlying hardware memory (Technology-alt). GPT-4 has over 1 trillion parameters, requiring on the order of several terabytes of memory, while NVIDIA H100 offers only 80 GB of memory. Since 2020, the per-LLM memory requirement has exceeded the memory offering per graphics processing unit (GPU), shifting us from a compute-limited era to a memory- and bandwidth-limited one. This widening gap, known as the "memory wall," forces data centers to scale the number of GPUs rather than relying on improvements in per-device capability, driving up energy and infrastructure costs. One fundamental reason memory scaling has stagnated is that dominant commercial memory technologies, DRAM and SRAM, are both volatile and single-bit. Volatile means each memory cell requires continuous power to preserve its stored value; if power is removed, the information is lost, which implies constant refresh operations in DRAM and static biasing in SRAM. Single-bit means that every physical cell stores only one binary value. In this article, we argue that a shift to non-volatile, multi-bit memory technologies would break both constraints: non-volatility would eliminate refresh energy and reduce standby power, while multi-bit storage would allow each cell to encode multiple discrete levels, notably increasing density per unit area. Together, these properties offer a path toward reducing the "memory wall" between compute and memory.

To make matters worse, historical microprocessor trend data show that the traditional scaling of von Neumann digital CMOS systems is saturating. While transistor counts per microprocessor have continued their exponential climb from just 4,000–8,000 transistors in the 1970s to well over 100 million by the mid-2000s and beyond 10 billion in 2020 (projected to reach 1 trillion by the first half of the 2030s), these increases no longer deliver proportional performance benefits (Liu and Wong 2024) (Rupp 2025). Single-thread performance, measured using a standard benchmark of per-core integer throughput, rose nearly 1000× between the late 1980s and 2010 but has remained essentially flat since then; likewise, clock frequencies increased steadily from a few MHz to roughly 3–4 GHz by the early 2000s but have not improved in nearly two decades, while chip power consumption climbed from only a few watts in the 1980s to 100–200 W in modern processors (Rupp 2025).

Koomey's law is the empirical observation that the energy efficiency of computing has doubled roughly every 1.57 years. Indeed, energy-per-operation data show that the energy required for a single useful computation has fallen by more than 12 orders of magnitude since the ENIAC era, going from about 100 J/operation in the 1940s, to microjoules in early personal computers (PCs), to nanojoules in systems such as the Pentium III and Sony PS3, and finally to the picojoule regime in modern accelerators like NVIDIA's P100 GPU and Google's tensor processing unit (TPU) (Marinella and Agarwal 2019). However, this trajectory is flattening as state-of-the-art systems approach the CMOS energy floor, typically estimated around 100 fJ/operation (Marinella and Agarwal 2019; Ho et al. 2023). As progress stalls, the exponential trend described by Koomey's law might not hold for long, and further improvement will require fundamentally new devices and materials rather than architectural optimization alone.

> "As classical computing systems approach the CMOS limit, further improvement will require fundamentally new devices and materials, as well as new ways to densely connect them."

These constraints force industry to rely on multicore parallelism, with logical core counts increasing from one core to 16, 32, or more cores in recent designs (Rupp 2025). This requires algorithms to shift from older, single-core libraries to interface with highly-distributed architectures in order to realize the increasing floating-point operations per watt seen over the past decade (Rupp 2013). Additionally, parallelism is still inherently limited by memory bandwidth and the von Neumann bottleneck. In this article, we will explore a new class of hardware known as neuromorphic computing that significantly differs from highly-parallel GPU architectures though combining digital, analog, and mixed-signal representations with dynamical systems, sparse communication, and memory-centric approaches at ultra-low power.

**Neuromorphic Computing**

Neuromorphic processors perform their core computations with a network of interconnected neural units and synapses. Neural units are physical circuits that receive and communicate electrical signals, placed in an arrangement akin to neurons in the brain, while synapses are learnable, weighted connections that modulate the strength of signals that are passed between these circuits. Mathematically, neural units are dynamical systems, with several internal states, each expressed as a differential equation, evolving according to their past states, external stimuli, and synapse-modulated inputs arriving from other neural units. This perspective naturally motivates compute-near-memory (CNM) architectures, in which computation occurs physically close to where information is stored, reducing data movement between memory and processor. In neuromorphic systems, CNM manifests as local dendritic, somatic, and synaptic computations occurring adjacent to the synaptic parameters, often stored in tiled SRAM across the chip.

Recently, a growing body of neuromorphic architectures have shifted from CNM to CIM, where memory elements directly participate in computation. Building on Carver Mead's original vision, neuromorphic circuits use the inherent physics of analog devices, such as capacitors, transconductance amplifiers, and current-mode circuits, to implement filtering, integration, and nonlinear dynamics at low precision with far fewer transistors than digital equivalents (e.g., a two-primitive analog low-pass filter versus dozens of digital logic blocks) (Mead 1989; Boahen 2017). In contrast, communication between neuromorphic circuits remains digital and spike-based, providing robustness and low energy per event. The devices responsible for implementing CIM are emerging non-volatile memories (eNVMs), such as resistive random-access memory (RRAM), phase change memory (PCM), spin-transfer torque magnetoresistive random-access memory (STT-MRAM), and ferroelectric memory (FeM), representing synaptic elements that both store weights without power and perform local computation such as multiplication and temporal filtering. Commercial eNVMs today are predominantly single-bit, with the aforementioned eNVMs offered by TSMC, STMicroelectronics, Intel, and GlobalFoundries, but their non-volatility and physical compute capabilities make them central to the future of CIM-based neuromorphic architectures (Yu et al. 2021).

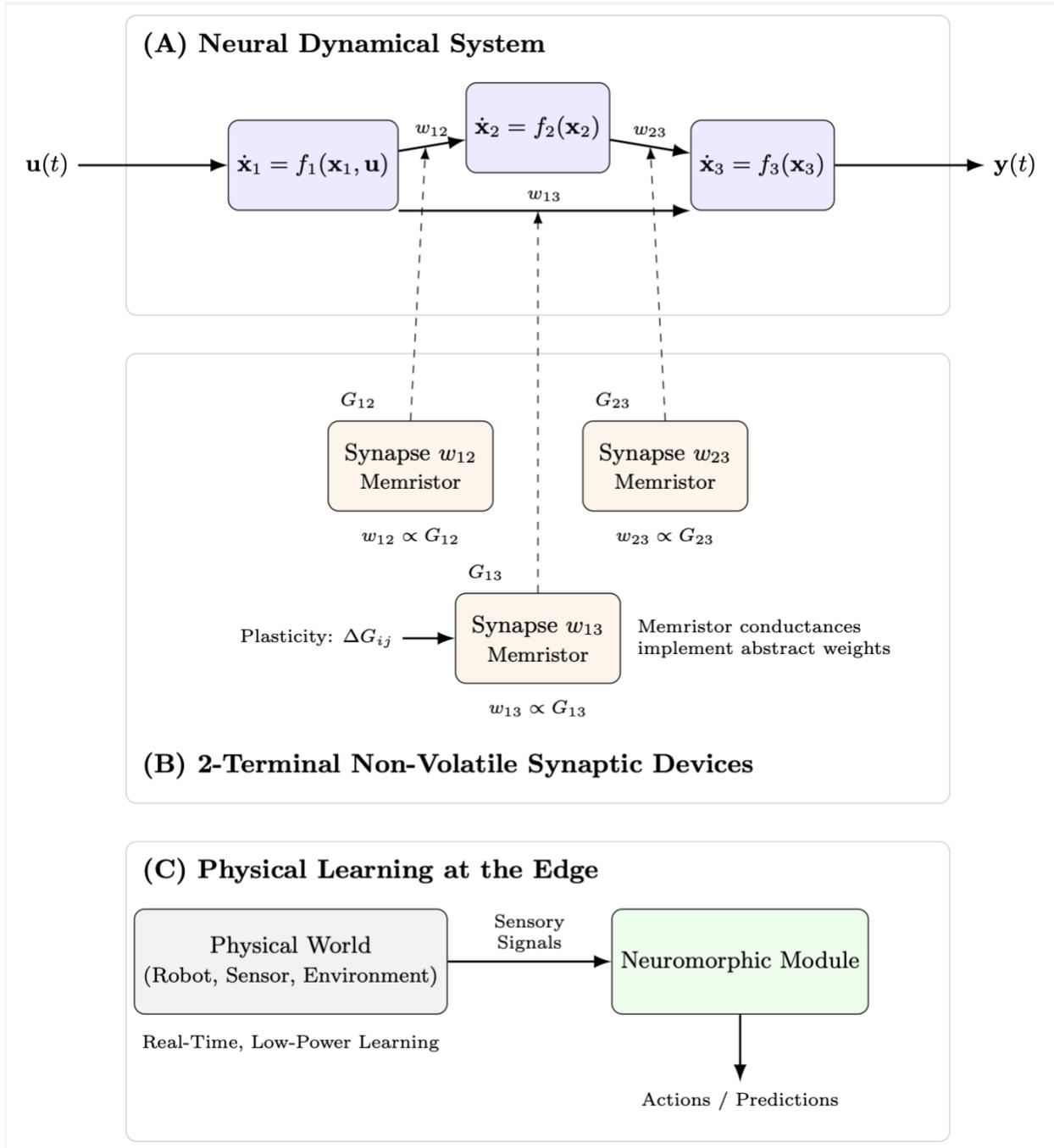

**Figure 2:** A near-future example of a memristor-based neuromorphic module that offers adaptive filtering, integration, differentiation, and Fourier analysis.

A central promise of neuromorphic computing is its ability to address the core limitation of classical CMOS systems: unfavorable energy scaling dominated by communication costs. Recall that moving data between physically separated memory and processing units in conventional von

Neumann architectures scales roughly with the cube of the system size, quickly overwhelming any gains from denser logic or faster arithmetic units. Neuromorphic architectures address this issue by restructuring where and how information flows. The first strategy is CNM, which reduces energy scaling to $O(N^2)$ by distributing small SRAM tiles across the chip so that neural updates occur physically close to stored parameters. The second approach employs 2D crossbar arrays of eNVMs, maintaining quadratic scaling but achieving reductions in area and latency by performing CIM operations directly within the memory fabric. A third pathway, increasingly explored in research systems, uses 3D-stacked eNVMs to push scaling even lower, approaching $O(N^{3/2})$, provided that sparse communication patterns limit the thermal load associated with vertical integration (Boahen 2022). Despite these improvements, none of the existing approaches achieve a per-neuron power consumption comparable to the human brain, leaving open an important problem in neuromorphic architecture. Achieving such scaling will require careful co-design across devices, circuit topology, and algorithms, combining optimized memory technologies with highly sparse spatial and temporal communication codes.

The 3D stacked processor architecture illustrated in Figure 1 (Right) is an ideal hybrid neuromorphic chip in which processing, memory, and communication are vertically integrated into a single tightly coupled hierarchy. At the bottom, conventional Si-CMOS processor units handle robust digital control, followed by a layer of CNM, and then higher layers containing primary memory arrays and CIM units fabricated in the back end of line (BEOL), the upper metal layers added after transistor fabrication to allow dense, low-cost integration of memory on top of CMOS logic without disturbing the underlying circuits.

The road to commercialization for neuromorphic chips involves weighing decisions about signal type, memory/compute integration, off-chip versus on-chip learning, and application focus (Muir and Sheik 2025). Current neuromorphic implementations span a spectrum from digital architectures close to commercialization to mixed-signal/analog and crossbar-based CIM systems that remain further from maturity. Digital neuromorphic chips offer simplicity and compatibility with advanced CMOS nodes, but they inherit the high switching and communication energy of binary logic. Moving toward mixed-signal and analog circuits can reduce power per operation at low precision by exploiting continuous device physics, but this comes with challenges such as mismatch, drift, and the need for more specialized fabrication steps. A similar progression is seen in the compute/memory architecture dimension. CNM architectures are closer to commercialization because they use conventional memory blocks and CMOS-friendly layouts. Further from commercial deployment are CIM architectures, typically built around 2D or 3D eNVM crossbars. Though CIM minimizes data movement and maximizes density, it faces unresolved device-level issues such as conductance variability, non-ideal analog behavior, and challenges integrating eNVM materials into standard CMOS workflows.

Table 1 summarizes a representative set of neuromorphic processors alongside the NVIDIA H100 SXM GPU. IBM TrueNorth was a pioneering neuromorphic chip unveiled in 2014 as part of DARPA's SyNAPSE, a research program which aimed to develop a computer with form, function, and architecture similar to the mammalian brain; this chip contains 1 million digital neurons and 256 million synapses, offering a 26 giga-synaptic operations per second (GSOPS) throughput, which measures how many tasks a processor can complete per unit of time, at a footprint of only 65 mW (Akopyan et al. 2015). TrueNorth has reached end-of-life status but has been followed up by two mixed-signal neuromorphic chips of note, BrainScaleS-2 and HERMES. The former is a research platform developed at Heidelberg University which, though unable to outperform TrueNorth in the metrics listed in Table 1, represents an important step in realizing analog neuromorphic hardware by offering a full-custom analog core with a synaptic crossbar, neuron circuits, and analog parameter storage (Pehle et al. 2022). Similarly, HERMES is IBM's most recent research chip focused on analog CIM, leveraging a crossbar array of PCM-based unit cells to build a 64-core neuromorphic processor with a throughput of up to 63.1 tera-operations per second (TOPS) (Le Gallo et al. 2023).

Intel Loihi 2 and SpiNNaker2 represent the most advanced processors in the neuromorphic space, the latter of which is commercially available by SpiNNcloud Systems. Both offer highly competitive energy efficiency at 16 TOPS/W and 6.4 TOPS/W, respectively, exceeding the 5.65 TOPS/W of the NVIDIA H100 (Mayr et al. 2019; Intel 2021; NVIDIA 2022). Today, digital spike-based neuromorphic processors such as Loihi 2 and SpiNNaker2 are primarily bottlenecked by software, training, and application alignment. Spiking neural networks (SNNs) remain difficult to train and program, toolchains are immature compared to deep learning stacks, and commercially scalable workloads that naturally exploit spikes are still limited. In contrast, mixed-signal compute-in-memory systems like HERMES are limited mainly by device- and circuit-level issues, including PCM variability, limited weight density, and the need for tight hardware-algorithm co-design, with ADC/DAC energy and software maturity playing a secondary role.

The application focus of these neuromorphic systems distinguishes between edge/online, which demand low-power, real-time inference, and data-center/offline that rely on large-scale compute infrastructure. There is growing interest in applying neuromorphic systems to memory-bound operations in transformer inference, but rack-scale integration of neuromorphic inference chips alongside NVIDIA GPUs is unlikely to be realized in the near-term since the current neuromorphic ecosystem lacks massive hardware, software, and batching capabilities required to integrate into data centers. Instead, neuromorphic technologies today align far more naturally with low-power edge computing or embodied robotics. Growing ecosystems with multimodal sensor inputs, specifically dynamic vision sensors (DVS), RGB cameras, and inertial measurement units (IMU), present an opportunity for neuromorphic solutions to navigation tasks, such as optical flow, pose estimation, and depth estimation. Recent demonstrations include a fully neuromorphic vision-to-control pipeline that uses Intel's Loihi for controlling an insect-size drone and a SpiNNaker-based system based on ant navigation for identifying routes through vegetation (Zhu et al. 2023; Paredes-Vallés et al. 2024). While Loihi and SpiNNaker offer on-chip training, most neuromorphic processors still depend on off-chip learning, where models are trained on GPUs or classical hardware and then deployed onto neuromorphic chips for inference. This approach keeps chip design simpler and leverages mature machine learning toolchains.

| Processor | Signal Type | Memory | Technology [nm] | Area [mm$^2$] | Energy Efficiency | Throughput | Power [W] |
|---|---|---|---|---|---|---|---|
| Loihi 2 | Digital | SRAM | 7 | 31 | 15 TOPS/W | 33.9 TOPS | 2.26 |
| SpiNNaker2 | Digital | SRAM | 22 | 8.76 | 6.4 TOPS/W | 4.6 TOPS | 0.72 |
| TrueNorth | Digital | SRAM | 28 | 430 | 400 GSOPS/W | 26 GSOPS | 0.065 |
| BrainScaleS-2 | Mixed | SRAM | 65 | 32 | 9.29 GOPS/W | 52 GOPS | 5.6 |
| HERMES | Mixed | PCM | 14 | 144 | 2.48/9.76 TOPS/W | 16.1/63.1 TOPS | 6.49 |
| H100 SXM | Digital | HBM3 | 5 | 814 | 5.65 TOPS/W | 3958 TOPS | 700 |

**Table 1:** Comparison of representative digital/mixed-signal and compute-near-memory/compute-in-memory architectures. For HERMES, energy efficiency and throughput is provided for four-phase (high-precision) / one-phase (low-precision) operational read mode. Source: (Akopyan et al. 2015; Mayr et al. 2019; Intel 2021; NVIDIA 2022; Pehle et al. 2022; Le Gallo et al. 2023).

Taken together, these trends indicate that the most promising near-term deployment space for neuromorphic processors is real-time, low-power inference with off-chip learning, particularly in domains such as edge computing, internet of things (IoT), autonomous sensing, and mobile or embodied robotics. These applications benefit directly from neuromorphic efficiency while avoiding the infrastructure and precision constraints that currently limit neuromorphic adoption in large-scale training or cloud settings. Figure 2 illustrates how these advantages manifest in a near-future neuromorphic system. Specifically, Figure 2 (A) depicts the system as a graph of interacting dynamical units connected by weighted synapses, while Figure 2 (B) shows how those abstract synaptic weights can be physically realized using two-terminal devices known as a memristors, whose conductances encode and maintain synaptic values with minimal energy. Figure 2 (C) proposes that the resulting neuromorphic module processes continuous sensory streams, performing adaptive filtering, integration, differentiation, or simple spectral transforms directly in hardware, making it well-matched for physical tasks in robotics and sensing environments.

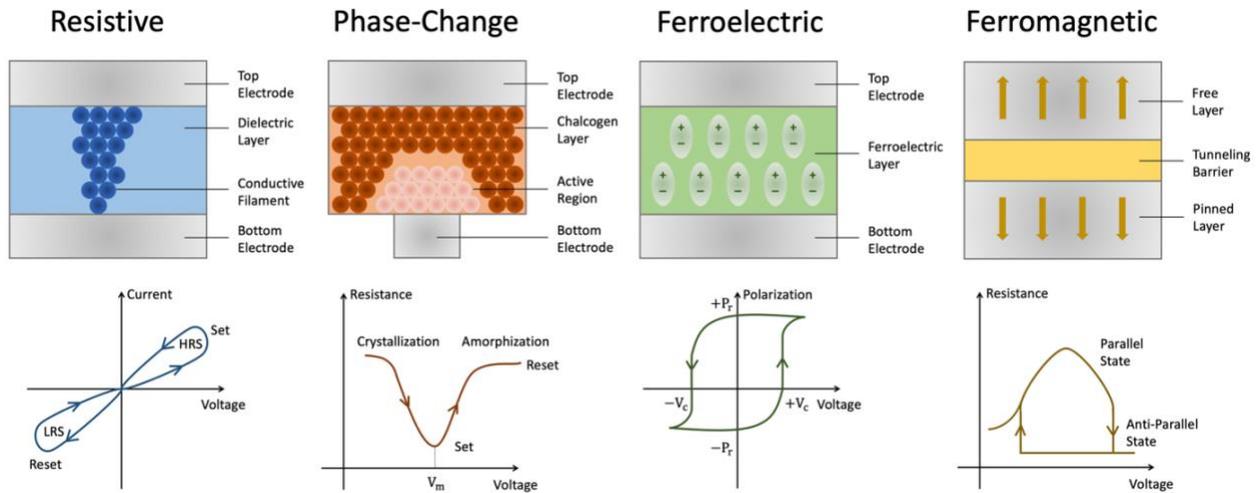

**Figure 3:** An overview of the four typical structures for emerging non-volatile memory. Source: (Liu et al. 2023).

**The Non-Volatile Memory Landscape**

Two core considerations are which eNVM device to select for a neuromorphic processor and how the resulting CIM would compare to existing memory technologies. Addressing the latter consideration, modern memory technologies fall into two broad categories: random-access memory (RAM) and flash memory, both of which store information using charge. SRAM and DRAM are the dominant RAM technologies. SRAM uses cross-coupled inverters to maintain fast, stable logic states at the cost of high area per bit, while DRAM stores charge on a capacitor and must be refreshed every few milliseconds, making it dense but inherently volatile and energy-hungry (Zahoor et al. 2020). Flash memory, although non-volatile, relies on charge stored in floating gates, resulting in slow programming, high operating voltages, and poor write endurance (Rehman et al. 2020). All three technologies face severe scaling limits below ~10 nm and, as previously discussed, remain physically separate from compute units, offering no relief from the von Neumann bottleneck. These limitations have motivated the development of eNVMs that aim to combine the speed of SRAM, density of DRAM/flash, and non-volatility of flash while providing multi-bit or analog conductance states suitable for CIM.

We now turn our attention to the former consideration by discussing the four main classes of eNVM devices. As illustrated in Figure 3, eNVMs are generally grouped into resistive, phase-change, ferroelectric, and ferromagnetic families. RRAM switches between high- and low-resistance states by forming or rupturing conductive filaments (Lanza et al. 2019). PCM exploits the large electrical contrast between crystalline and amorphous phases of chalcogenide materials, toggled using heat-generating electrical pulses (Burr et al. 2010). STT-MRAM and related ferromagnetic devices encode bits through the relative magnetization of two layers, parallel for low resistance and antiparallel for high resistance (Ielmini and Wong 2018). Ferroelectric memories, including ferroelectric random-access memory (FeRAM) and the more recently commercialized ferroelectric field-effect transistor (FeFET), use voltage-controlled polarization states in ferroelectric films to achieve fast, non-volatile operation. A particularly promising ferroelectric device is the AlScN-based ferrodiode (FeD), a two-terminal ferroelectric CIM device with low fabrication complexity that has demonstrated 5-bit operation, 10 fJ/bit switching energy, ultra-fast switching speed, and as little as a 3 V operating voltage (Liu et al. 2021; Liu et al. 2022; Kim et al. 2024; Sarkar et al. 2025).

Although a FeD shares the same metal-ferroelectric-metal (MFM) structure as a similar device known as a ferroelectric tunnel junction (FTJ), the way it conducts current is fundamentally different. In a FeD, the ferroelectric layer is too thick for electrons to move by direct tunneling. Instead, current flows through polarization-dependent leakage processes, such as defect-assisted tunneling or Poole-Frenkel emission, which makes the current highly sensitive to the height of the interfacial Schottky barrier (Sarkar et al. 2025). This results in a self-rectifying I–V curve, meaning the device naturally allows current to flow much more easily in one direction than the other. In large crossbar memory arrays, each cell sits at the intersection of a word line and a bit line, and unwanted "sneak-path" currents can flow through unselected cells during read or write operations. Conventional designs require a dedicated selector device, typically a transistor, to block these parasitic currents. The self-rectifying behavior of a FeD inherently suppresses sneak-paths, allowing the memory cell to operate without a selector, simplifying the structure from a 1T1C cell to a compact 1R cell.

Many of the advantages of the aforementioned FeDs are thanks to aluminum-scandium nitride (AlScN), which has emerged as a leading ferroelectric material for generation non-volatile memory because it combines several exceptional properties rarely found in a single system (Fichtner et al. 2019). It exhibits the highest known remnant polarization ($P_r$), which is the amount of polarization that remains after an external electric field is removed, enabling strong, stable memory states. AlScN is also uniphasic, meaning it maintains a single crystallographic phase across operating conditions, which contributes to its extraordinary thermal stability and highest known Curie temperature ($T_c$) among thin-film ferroelectrics. In addition, it offers the lowest permittivity ($k$) together with the highest $P_r$. This combination enables the low switching energy, fast switching speed, and precise multi-bit operation through partial polarization control seen in FeDs. Crucially, AlScN is fully compatible with BEOL semiconductor processing, allowing ferroelectric memory elements to be integrated directly above CMOS logic without thermal or material conflicts, an essential requirement for scalable in-memory computing. Due to this unique set of advantages, FeDs based on AlScN have already been integrated into a 2-kilobyte (128 × 128) stable crossbar array, as illustrated in Figure 4 (Han et al. 2025). Scaling FeDs to a 90-nm diameter with a SkyWater process could produce a 100-megabyte array in a similar footprint (Custom ASIC | Process Design Kit (PDK) & IP Design | SkyWater 2025).

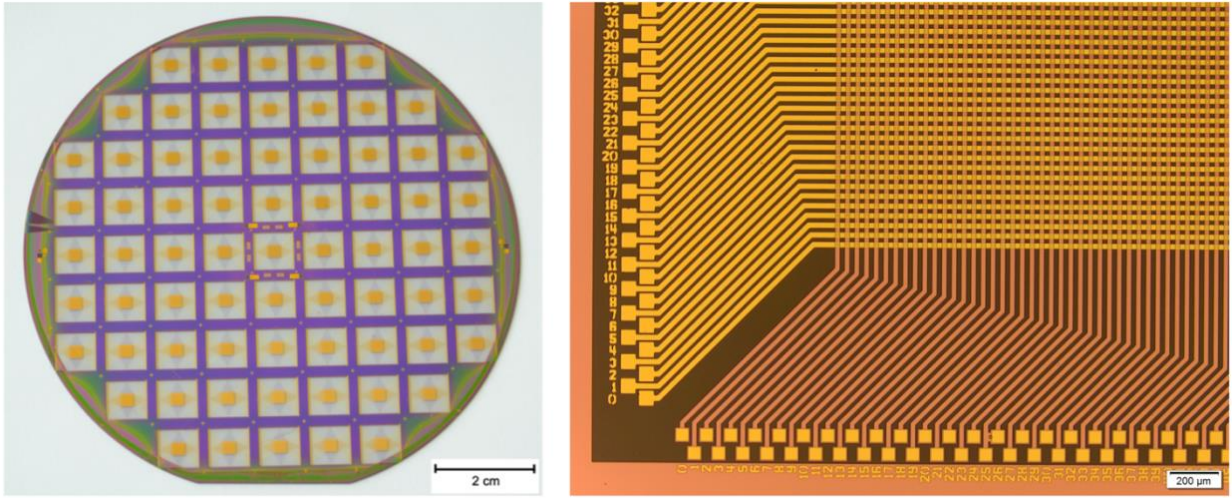

**Figure 4:** Demonstration of 2-kilobyte AlScN-based ferrodiode array. **(Left)** Camera image of 5 µm crossbar array on a 4-inch wafer. **(Right)** Microscope image of bottom-left section of 10 µm crossbar array. Source: (Han et al. 2025).

Two notable extensions of the FeD demonstrate its potential well beyond conventional edge computing. First, FeDs operate reliably across high-temperature, cryogenic, and high-radiation environments, making them uniquely suited for aerospace, deep-earth, nuclear, and extreme industrial settings, domains where traditional CMOS devices rapidly degrade (Pradhan et al. 2024; He et al. 2025; Song et al. 2025). In these environments, transistor density will not scale according to Moore's law, meaning that digital logic will grow only slowly or not at all (Pradhan et al. 2024). Because neuromorphic architectures leverage memory-rich, low-power computation, they offer one of the only viable pathways for achieving scalable computation under such harsh conditions. A second extension involves integrating FeDs with two-dimensional (2D) materials, which promise ultra-low-power switching and high energy efficiency. Combining AlScN with transition metal dichalcogenides (TMDCs), such as $MoS_2$, $WS_2$, or $WSe_2$, or with carbon nanotube (CNT) or hexagonal boron nitride (hBN) heterostructures creates ferroelectric-2D systems that naturally minimize parasitic capacitance and enable strong ferroelectric gating of atomically thin channels. Furthermore, monolithic three-dimensional (M3D) integration, an alternative to traditional through-silicon via technology that increases the density of stacked electronic components, has been shown for $MoS_2$ memtransistors (Ghosh et al. 2024). 2D/AlScN and $MoS_2$ memtransistor platforms have already been demonstrated in scalable form, highlighting its viability for future embedded memory technologies (Kim, Oh, et al. 2023; Schranghamer et al. 2025). Other exploration of 2D materials in computing devices have shown $WSe_2$ field-effect transistors (FETs) that accelerate a stochastic inference engine on medical image classification and a lobula giant movement detector (LGMD) neuron in locusts that is emulated via $MoS_2$ FETs (Jayachandran et al. 2020; Ravichandran et al. 2024).

Standard metrics for evaluating eNVMs include the ON/OFF ratio, relating to the number of distinguishable conductance/polarization states for multi-bit operation; write energy and write speed, quantifying the efficiency and speed of programming; read speed, governing inference throughput; endurance, the number of cycles a device can repeatedly switch without degradation; and retention, the length of time a state can be preserved without power. AlScN-based FeDs exhibit ON/OFF ratios of $10^2 - 10^3$, write energy of < 50 fJ/bit, read/write speed of < 10 ns, endurance of $> 10^9$, retention of $> 10$ years, and multi-bit performance at 3–5 bits, with all metrics but endurance outperforming RRAM, PCM, STT-MRAM, DRAM, and flash, making them one of the most promising candidates for next-generation neuromorphic and CIM systems (Hu et al. 2025). Reliability and reproducibility concerns, such as conductance drift, finite endurance, retention loss, and temperature dependence, are rarely addressed at the device level alone. Instead, they are mitigated through a combination of periodic calibration and refresh, closed-loop training that adapts to device evolution, and algorithm-level robustness, including noise-aware training or error-tolerant model design. This cross-layer approach allows non-idealities to be absorbed or compensated at higher abstraction levels and, in some cases, can even be leveraged for new computational paradigms (Ravichandran et al. 2024).

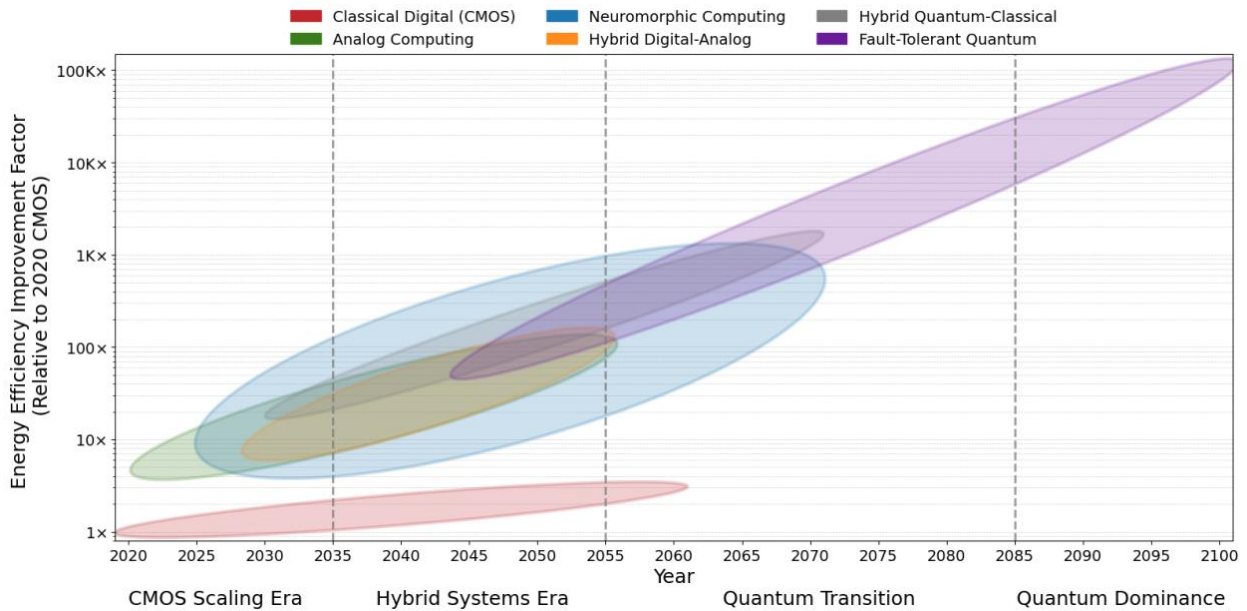

**Figure 5:** Approximate predictions from trends/theoretical estimates of the energy efficiency improvement factor (relative to 28-nm node CMOS from 2020) of various high-performance computing technologies (Traversi et al. 2024).

**The Future of High-Performance Computing**

Predicting the hardware platforms that will drive the next century of high-performance computing is essential for guiding both research and investment. Such predictions must remain realistic. That is, they must balance the promise of emerging substrates with the inertia, cost structures, and physical constraints of existing semiconductor manufacturing. Today, six major hardware classes shape the technological horizon. First, we have classical digital CMOS, which remains dominant due to its robustness and manufacturability, but as mentioned throughout this article, it is fundamentally limited by switching energy, interconnect costs, and the breakdown of Moore's law. Analog computing leverages continuous device physics to perform low-precision arithmetic with extremely high energy efficiency, though it suffers from noise, mismatch, and calibration challenges, while hybrid digital-analog systems blend efficient analog primitives with digital robustness to overcome the limitations of either approach alone, representing our second and third hardware classes. Fourth, neuromorphic computing combines all three of the previous classes with dynamical systems, sparse communication, and memory-centric architectures to emulate aspects of neural processing with ultra-low power. Fifth and sixth are hybrid quantum-classical accelerators that couple near-term quantum processors with conventional central processing units (CPUs) or GPUs for specific algorithmic subroutines, as well as fault-tolerant quantum computers, which represent the long-term ambition of scalable, error-corrected quantum processing.

Figure 5 provides an approximate comparison of these technologies by projecting their potential energy efficiency improvement factors relative to 2020 CMOS per the authors, as well as various estimates in literature. The trends suggest that over the next 10–20 years, the most economically impactful advances will come from hybrid classical approaches, particularly those rooted in analog and CIM neuromorphic architectures built from emerging materials and devices. These systems can plausibly achieve 10–100× improvements in energy efficiency while remaining compatible with commercial semiconductor processes. In contrast, the transformative advantages of quantum computing, especially fault-tolerant quantum systems, are unlikely to materialize until the second half of the century, when energy efficiency improvements could grow to $10^4 - 10^6 \times$ for narrow classes of algorithms but only after extensive breakthroughs in error correction at architecture level and coherence, fabrication, and scaling at the qubit level (Fellous-Asiani et al. 2023; Trochatos et al. 2025).

> "Over the next 10–20 years, hybrid classical approaches like neuromorphic computing are likely to make a large economic impact."

A final point of comparison often invoked in the neuromorphic community is the human brain, which performs continuous, real-time inference, control, and learning with remarkable energy efficiency at roughly 25 W to sustain approximately $10^{11}$ neurons and $10^{14} - 10^{15}$ synapses (Boahen 2022). This is a compelling heuristic benchmark, but it is not a strict target. That is, achieving similar neuron/synapse counts at comparable power does not guarantee equivalent functionality, expressivity, or cognitive capacity. Nor do we yet understand which biological principles are essential versus incidental for intelligence. Many layers of complexity, ranging from dendritic computation and neuromodulation to molecular signaling and lifelong plasticity, separate today's neuromorphic devices from biological neural systems. These unknowns ensure that the relationship between neuromorphic hardware and the brain will remain a rich and expansive space of research for decades.